\documentclass[runningheads]{llncs}

\usepackage[T1]{fontenc}
\usepackage{graphicx}
\usepackage{amsmath,amsfonts}
\usepackage{algorithmic}
\usepackage{algorithm}
\usepackage{array}
\usepackage[caption=false,font=normalsize,labelfont=sf,textfont=sf]{subfig}
\usepackage{textcomp}
\usepackage{stfloats}
\usepackage{url}
\usepackage{verbatim}
\usepackage{graphicx}
\usepackage{cite}

\usepackage{amsmath,amssymb,amsfonts}
\usepackage[algo2e,ruled,linesnumbered]{algorithm2e} 
\usepackage{array}
\usepackage[caption=false,font=normalsize,labelfont=sf,textfont=sf]{subfig}
\usepackage{textcomp}
\usepackage{stfloats}
\usepackage{url}
\usepackage{verbatim}
\usepackage{graphicx}
\usepackage{wrapfig}
\usepackage{soul}
\usepackage{color}
\usepackage{multirow}
\usepackage{makecell}
\usepackage[normalem]{ulem}

\usepackage[misc]{ifsym}
\usepackage{bbding}

\usepackage{amssymb}
\usepackage{pifont}

\usepackage{cite}
\usepackage{graphicx}
\usepackage{textcomp}
\usepackage{xcolor}

\usepackage{boxedminipage}
\usepackage{color}
\usepackage{hyperref}
\usepackage{wrapfig}

\usepackage{bbm}
\usepackage{graphicx}
\usepackage{mathtools}
\usepackage{booktabs} 
\usepackage{bbding}

\usepackage{verbatim}
\usepackage{multirow}

\usepackage{orcidlink}

\usepackage{amsmath}
\usepackage{soul}

\usepackage{comment}
\usepackage{textcomp}

\usepackage{comment}
\usepackage{multicol}

\newcommand{\squishlist}{
   \begin{list}{$\bullet$}
    { \setlength{\itemsep}{0pt}      \setlength{\parsep}{0pt}
      \setlength{\topsep}{3pt}       \setlength{\partopsep}{0pt}
      \setlength{\listparindent}{-2pt}
      \setlength{\itemindent}{-5pt}
      \setlength{\leftmargin}{1em} \setlength{\labelwidth}{0em}
      \setlength{\labelsep}{0.5em} } }

\newcommand{\squishend}{
    \end{list}  }
    
\usepackage{multirow}  
\usepackage{comment}
\usepackage{listings}
\usepackage{xcolor}
\usepackage{footnote}
\usepackage{threeparttable}
\usepackage{paralist}

\usepackage{amssymb}  
\usepackage{booktabs}

\usepackage{xcolor}
\usepackage[normalem]{ulem} 

\usepackage{tikz}
\usetikzlibrary{calc}

\usepackage{tikz}

\newcommand*\blackcircled[1]{\tikz[baseline=(char.base)]{
        \node[shape=circle,fill={rgb,255:red,0;green,0;blue,0}, text=white, font=\small, inner sep=0.6pt] (char) {#1};}}

\newcommand*\blackcircledempty[1]{\tikz[baseline=(char.base)]{
        \node[shape=circle, text={rgb,255:red,0;green,0;blue,0}, font=\small, draw={rgb,255:red,0;green,0;blue,0},inner sep=0.5pt] (char) {#1};}}

\begin{document}
\title{Accelerating Mini-batch HGNN Training by Reducing CUDA Kernels}
\titlerunning{HiFuse}
%
\author{Meng Wu\inst{1,2}  \and Jingkai Qiu\inst{3}  \and Mingyu Yan\inst{1,2,\textsuperscript{\ding{41}}} \and Wenming Li\inst{1,2}  \and \\Yang Zhang\inst{4}\and Zhimin Zhang\inst{1,2}  \and Xiaochun Ye\inst{1,2} \and Dongrui Fan\inst{1,2}}
\authorrunning{M. Wu et al.}

%
\institute{SKLP, Institute of Computing Technology, Chinese Academy of Sciences 
\and University of Chinese Academy of Sciences 
\and Shanghai Tech University, Shanghai, China  
\and Yancheng Zhongke High Thoughput Computing Research Institute Co., LTD, Jiangsu, China \\
\email{\{wumeng, yanmingyu, liwenming, zzm, yexiaochun, fandr\}@ict.ac.cn, qiujk@shanghaitech.edu.cn, and zhangyang@smart-core.cn}}

\maketitle              
\begin{abstract}

Heterogeneous graph neural networks (HGNNs) are essential for capturing the structure and semantic information in heterogeneous graphs. However, existing GPU-based solutions, such as PyTorch Geometric, suffer from low GPU utilization due to numerous short-execution-time and memory-bound CUDA kernels during HGNN training.

To address this issue, we introduce HiFuse, an enhancement for PyTorch Geometric designed to accelerate mini-batch HGNN training on CPU-GPU systems. From the data perspective, we reorganize and merge multiple smaller vertex feature matrices into larger ones, enabling a single kernel to process larger data chunks. This efficiently exploits data locality, reduces the kernel launch overhead, and improves overall GPU utilization. From the workflow perspective, we sophisticatedly offload the construction of semantic graphs from GPU to CPU to reduce the number of CUDA kernels. To meet the parallelism requirements on CPU and ensure seamless execution between CPU and GPU, we employ parallelization techniques including multi-threading and asynchronous pipeline. This allows different stages of the process to overlap, enhancing GPU utilization and reducing end-to-end execution latency, leading to a more efficient and balanced use of computational resources. Through extensive experiments, HiFuse demonstrates an average 2.38$\times$ speedup compared to a state-of-the-art solution.


\end{abstract}
\section{Introduction} \label{sec:intro}

Heterogeneous graph neural networks (HGNNs) have found extensive application in processing graph data, adept at capturing complex relationships among diverse entities in real-world networks. HGNNs are a class of graph neural network (GNN) models tailored to handle heterogeneous graphs, where vertices and edges can belong to different types and carry varying semantics, such as knowledge graphs~\cite{oh2018knowledge, zhang2019iteratively, chen2017task}, social networks~\cite{yasunaga2019scisummnet,zheng2020clustering}, and others. They have achieved exceptional prediction accuracy in various critical fields~\cite{hgnn_survey_tangjie,hgnn_survey_hanjiawei,hgnn_survey_shichuan,hgnn_survey_shiruipan}, including recommendation systems~\cite{li2022disentangled,fan2019metapath}, medical analysis~\cite{luo2021imas}, knowledge inference~\cite{CompGCN,M2GNN}, malicious account detection~\cite{liu2018heterogeneous}, and information retrieval~\cite{mao2020item}.

The execution of an HGNN layer typically involves four major stages~\cite{HGNN_characterization,HiHGNN,SeHGNN}: Semantic graph build stage partitions the original heterogeneous graph into several semantic graphs, each focusing on different types of vertices and edges; Feature projection stage transforms the feature vectors of vertices in each semantic graph using a multi-layer perceptron (MLP) or similar neural network; Neighbor aggregation stage aggregates features from neighboring vertices for each vertex within each semantic graph to capture relational information; Semantic fusion stage integrates semantic information across all semantic graphs by combining results from the neighbor aggregation stage to provide holistic vertex representations.
These stages are repeated across multiple layers, and the final vertex representations are utilized for downstream tasks.

To accelerate the execution of HGNNs, several GPU-based solutions have been developed. The need for this acceleration stems from the computational complexity of processing large-scale heterogeneous graph data~\cite{HGL,HiHGNN}. As real-world heterogeneous graphs grow in size and complexity, the CPU-based or GPU-based solutions for traditional GNNs become increasingly inefficient for HGNN training tasks.
Previous solutions~\cite{PyTorch_Geometric,DistDGLv2,HGL} leverage GPUs' parallel processing capabilities to meet the computational demands of HGNN training. For instance, PyG (PyTorch Geometric)~\cite{PyTorch_Geometric}, a widely used HGNN acceleration framework, significantly enhances efficiency and scalability in heterogeneous graph learning. It provides a unified framework for handling multiple vertex and edge types, optimized sampling methods, and type-specific operations, utilizing PyTorch's tensor operations and GPU acceleration. Compared to traditional neural network frameworks, PyG yields substantial performance improvements, making it a potent tool for efficiently executing HGNNs.

Previous GPU-based solutions for HGNN training suffer from low GPU utilization primarily due to the excessive number of semantic graphs, which generate numerous short-execution-time and memory-bound GPU kernels during both the neighbor aggregation and semantic graph build stages.
In the neighbor aggregation stage, a set of short-execution-time CUDA kernels is assigned to process each semantic graph. As the number of semantic graphs increases, so does the number of these kernels, exacerbating the overhead from frequent kernel launches~\cite{kernel_launch_overhead}. Moreover, these kernels are generally memory-bound due to their graph-topology-dependent program behavior~\cite{HiHGNN, HGNN_characterization,HyGCN,GCN_Charaterization_CAL}, further contributing to low GPU utilization.
Similarly, in the semantic graph build stage, the CUDA kernels used have short execution time, and their number increases with the number of semantic graphs. This results in significant overhead from frequent kernel launches~\cite{kernel_launch_overhead}, compounded by the fact that these kernels are generally memory-bound due to their graph-topology-dependent program behavior~\cite{MetaNMP}, leading to under-utilization of GPU compute resources.
In summary, each kernel's brief execution time and the rapid succession of launches create substantial overhead, limiting GPU to achieve high utilization and efficient parallelism.

To address the issue of low GPU utilization in HGNN training, we introduce HiFuse, an enhancement for PyTorch Geometric aimed at accelerating mini-batch HGNN training on CPU-GPU systems through optimizations from both data and workflow perspectives. 

From the data perspective, we propose a method to reorganize and merge features of all semantic graphs to optimize memory access and reduce the number of kernels, improving GPU processing efficiency. This involves two key steps.

\squishlist

    \item Reorganization: Adjusting the layout of vertex features in semantic graphs to enhance memory access patterns. By grouping vertex features of the same type of vertices together and aligning them with GPU memory access patterns, we improve data locality in the neighbor aggregation stage. This ensures that computations for each semantic graph only need to access and efficiently reuse features of the same vertex type, thereby enhancing memory efficiency.

    \item Merging: Combining multiple smaller vertex feature matrices from semantic graphs into larger ones enables a single kernel to process larger data chunks, thereby reducing the number of kernels required for the neighbor aggregation stage. Utilizing a single kernel to handle the restructured data significantly decreases the overhead associated with kernel launches and increases overall data throughput on the GPU, enhancing GPU utilization.
    
\squishend

From the workflow perspective, we offload most of the semantic graph build stage from GPU to CPU, thereby avoiding the overhead of numerous small CUDA kernel launches and under-utilization of compute resources. The offloading part is control-intensive and only involves integer computation, which is better suited for CPU processing. To ensure this transition does not introduce bottlenecks, we employ two strategies.

\squishlist

    \item Parallelization: Implementing parallel processing on CPU to handle multiple semantic graph building concurrently using multi-threading techniques to efficiently utilize all available CPU cores.

    \item Asynchronous Pipeline: Introducing an asynchronous pipeline to overlap the execution of different stages across CPU and GPU. This overlapping ensures effective utilization of both CPU and GPU, minimizing idle times.

\squishend

By integrating these approaches, HiFuse not only improves GPU utilization but also reduces overall execution latency, making HGNNs more practical for large-scale, real-world applications. We summarize our contributions as follows.

\squishlist
    \item Performance Bottleneck Characterization: We quantitatively characterize the mini-batch training of HGNN, revealing that the state-of-the-art HGNN training framework suffers from low GPU utilization due to numerous short-execution-time and memory-bound kernels during the building and neighbor aggregation of excessive semantic graphs.

    \item Optimized Semantic Graph Processing: We propose a method to reorganize and merge vertex features across all semantic graphs, enabling a single CUDA kernel to efficiently utilize data locality and enhance GPU utilization.
   
    \item Enhanced Execution Workflow:  We offload most of the semantic graph build stage from GPU to CPU. By employing parallelization and pipeline techniques, we ensure seamless execution between CPU and GPU, reducing the number of kernels and improving overall performance.

    \item Comprehensive Evaluation: Through extensive experiments, we demonstrate the effectiveness of HiFuse, showing an average 2.38$\times$ speedup compared to a state-of-the-art framework.

\squishend

\section{Background} \label{sec:backgroud}

\subsection{Heterogeneous Graph} 

A heterogeneous graph is a type of graph that contains multiple types of vertices and edges, each representing different kinds of entities and relations~\cite{HG_survey}. Unlike homogeneous graphs, which have a single type of vertex and edge, heterogeneous graphs capture the complexity and diversity of real-world data. 
By incorporating rich semantic information through diverse vertex and edge types, heterogeneous graphs enable more nuanced analysis and learning~\cite{hgnn_survey_shiruipan,hgnn_survey_hanjiawei,hgnn_survey_shichuan,hgnn_survey_tangjie}.

Metapath is a crucial concept in heterogeneous graphs, defining a sequence of vertex types and edge types that outline a path through the graph. It captures complex structural patterns and semantic relationships between different types of entities, aiding in the understanding and definition of connectivity and interaction patterns within heterogeneous graphs.

Semantic graphs are subsets of the original graph that focus on specific types of vertices and relations, often derived based on metapaths. They are used to enhance the understanding of semantic relations and to facilitate more efficient and targeted computations in heterogeneous graphs.

\begin{figure}[!htbp]
    \vspace{-10pt}
    \centering
    \includegraphics[page=1, width=0.96\textwidth]{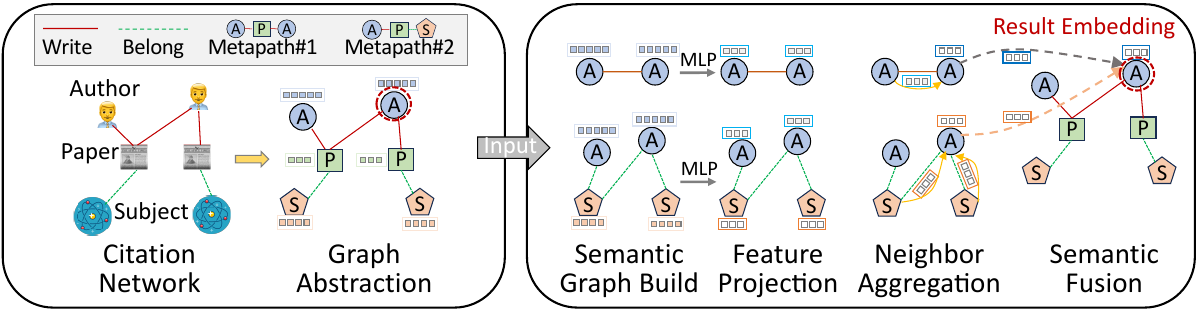}
    \vspace{-10pt}
    \caption{Illustrations for heterogeneous graph and HGNN.}
    \label{fig:hg_hgnn}
    \vspace{-10pt}
\end{figure}

An example of a heterogeneous graph is the ACM (Association for Computing Machinery) graph, as shown in Fig.~\ref{fig:hg_hgnn}. In this graph, different types of vertices represent various entities such as papers (P), authors (A), and subjects (S). The edges between these vertices denote different types of relations, such as an author writing a paper or a paper related to a subject. This rich structure allows the ACM graph to model the multifaceted interactions within the academic research community, capturing the complexity of real-world academic networks. By leveraging the diverse vertex and edge types, the ACM graph enables sophisticated analyses and insights, such as identifying influential authors and discovering emerging research subjects. Metapaths can be used to define paths through the ACM graph, such as ``Author - Paper - Subject'', which captures the sequence where an author writes a paper that is related to a subject.

\subsection{Heterogeneous Graph Neural Network}

HGNNs are a class of GNN models specifically designed to handle heterogeneous graphs~\cite{hgnn_survey_tangjie,hgnn_survey_shiruipan,hgnn_survey_shichuan}. They offer a powerful ability for processing and learning from these graphs by effectively capturing and integrating the complex relations among different types of entities. This capability is crucial for accurately modeling the multifaceted nature of real-world networks, such as academic networks, knowledge graphs, biological networks, and others~\cite{hgnn_survey_tangjie,hgnn_survey_shiruipan,hgnn_survey_shichuan}.
For example, in academic networks like the ACM graph, HGNNs can effectively model the relations between papers, authors, subjects, and other entities. They can distinguish between various types of connections, such as co-authorship, providing a more nuanced understanding of academic collaborations and research trends.

An HGNN layer typically involves four stages~\cite{HiHGNN,HGNN_characterization,SeHGNN}, as shown in Fig.~\ref{fig:hg_hgnn}.

\squishlist

    \item Semantic Graph Build: This stage splits the heterogeneous graph into multiple semantic graphs, each focusing on different types of vertices and edges. By isolating specific types of relations (metapaths), the model can better capture the unique interactions and dependencies within each subset of the graph.

    \item Feature Projection: In this stage, the feature vectors of vertices in each semantic graph are transformed using an MLP or a similar neural network. This transformation is essential for mapping the raw features into a latent space where they are more suitable for subsequent processing. It enhances the model's ability to learn from complex and high-dimensional data.

    \item Neighbor Aggregation: This stage involves aggregating features from neighboring vertices for each vertex within each semantic graph. By doing so, the model captures the relational information inherent in the graph structure. This aggregation process is crucial for understanding how information propagates through the network and how the local structure influences each vertex.

    \item Semantic Fusion: The final stage integrates semantic information across all semantic graphs by combining the results from the previous stage. This fusion process provides holistic vertex representations that incorporate information from all types of vertices and edges. These enriched representations are essential for accurately capturing the diverse relations in the heterogeneous graph.

\squishend

The final vertex representations generated by HGNNs can be used for various downstream tasks, such as vertex classification, where each vertex is assigned a label based on its features and relations; link prediction, which involves predicting the likelihood of a link between two vertices; and graph clustering, which groups vertices into clusters based on their similarities.

\subsection{HGNN Training}

\textbf{Full-batch and Mini-batch HGNN Training.}
HGNN training can be divided into two approaches: full-batch and mini-batch~\cite{GNN_DT_Survey}.

In full-batch training, the entire graph dataset is fed into the model as a single complete batch. This means that all vertices and edges participate in every forward and backward pass of the model. This approach is commonly used for small datasets or when maximizing the utilization of training data is crucial. However, it may encounter memory limitations and computational inefficiencies when dealing with large graph data.

On the other hand, mini-batch training divides the original graph dataset into multiple smaller batches, with each batch containing a subset of vertices and edges. These mini-batches are used for feature collection and model computation. This approach effectively handles large graph data by reducing the amount of data processed in each training iteration, thereby reducing memory consumption and improving computational efficiency.

Our work focuses on mini-batch training because it significantly improves efficiency and speed by optimizing hardware utilization and reducing memory consumption~\cite{GNN_DT_Survey}. Mini-batch training enhances convergence stability and supports scalable training on large datasets. It enables the effective use of vectorized operations and facilitates distributed computing.

\begin{figure}[!t]
    \centering
    \includegraphics[page=1, width=0.8\textwidth]{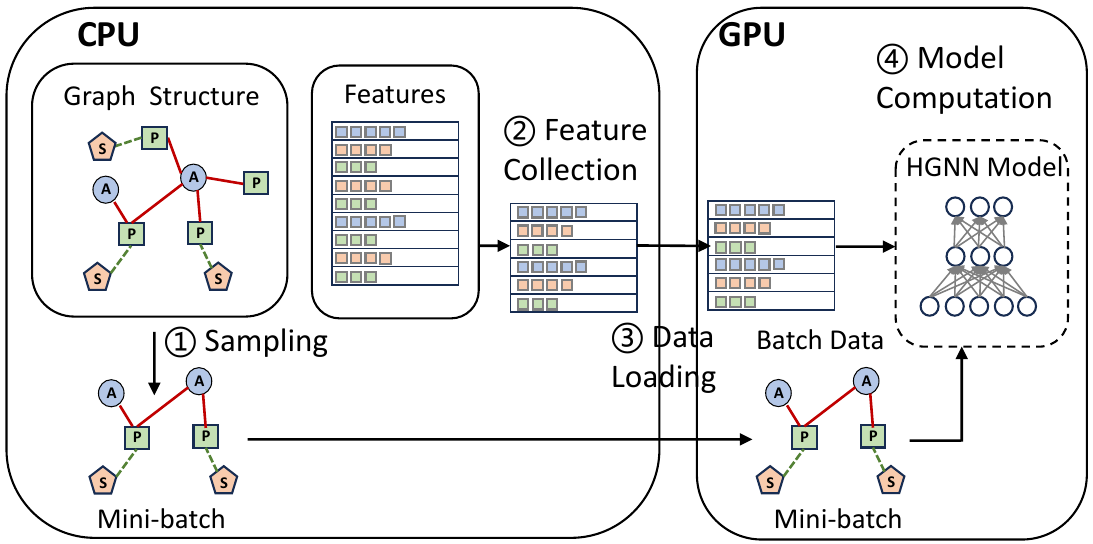}
    \vspace{-10pt}
    \caption{Workflow of mini-batch HGNN training.}
    \label{fig:workflow_hgnn_training}
    \vspace{-10pt}
\end{figure}

\textbf{Workflow of Mini-batch HGNN Training.}
Mini-batch HGNN training follows a structured workflow designed to efficiently process and learn from heterogeneous graph~\cite{PyTorch_Geometric}, as illustrated in Fig.~\ref{fig:workflow_hgnn_training}. \blackcircledempty{1} Sampling: mini-batches are sampled from the original graph on CPU. \blackcircledempty{2} Feature Collection: features specific to each mini-batch are then collected on CPU according to its topology. \blackcircledempty{3} Data Loading: the mini-batch and features are transferred from CPU to GPU. \blackcircledempty{4} Model Computation: these features undergo a forward pass on GPU through a series of stages, including semantic graph build, feature projection, neighbor aggregation, and semantic fusion. Then, the model performs a backward pass on GPU for gradient computation and parameter update.

\section{Motivation} \label{sec:motivation}

Previous GPU-based solutions encounter low GPU utilization due to numerous short-execution-time GPU kernels, primarily caused by the building and neighbor aggregation of a large number of semantic graphs. We demonstrate these inefficiencies using profiling results from PyG, the state-of-the-art HGNN training framework, running on a Linux server with an Intel Xeon Silver 4208 CPU and an NVIDIA T4 GPU. The profiling data, obtained via NVIDIA Nsight System and Compute, involves running PyG~\cite{PyTorch_Geometric} on the RGCN model with the AM dataset. Detailed experimental configurations are provided in Section~\ref{sec:Methodology}.

\begin{figure}[!t]
    \centering
    \includegraphics[page=1, width=0.88\textwidth]{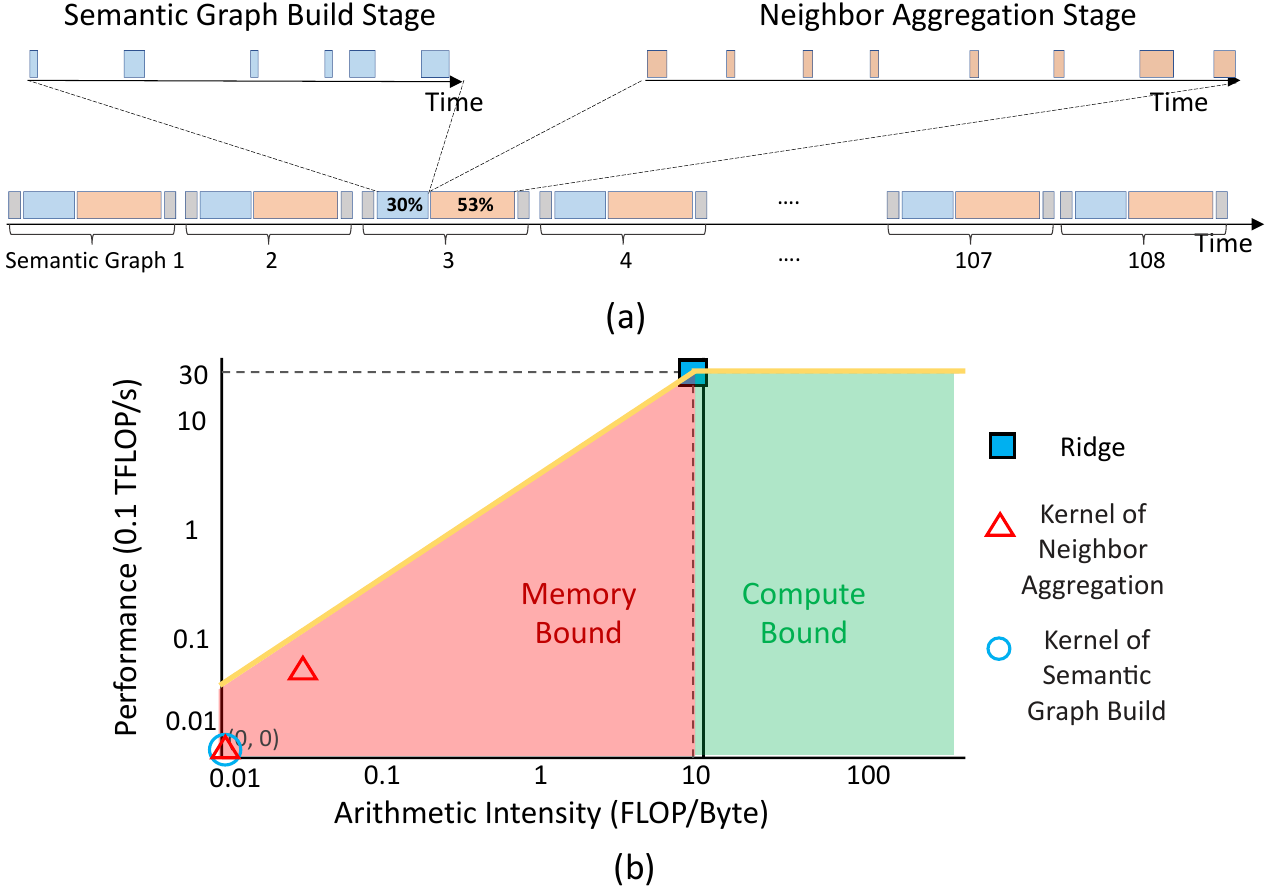}
    \vspace{-10pt}
    \caption{Frequent kernel launches of many short-execution-time and memory-bound kernels in semantic graph build and neighbor aggregation stages on RGCN model with AM dataset: (a) Partial timeline showing CUDA kernels' activity; (b) Roofline model of GPU FP32 performance showing CUDA kernels' execution bound.}
    \label{fig:motivation_overhead}
   \vspace{-20pt}
\end{figure}

\textbf{Inefficiency in the Semantic Graph Build Stage.} In the semantic graph build stage, CUDA kernels are employed with short execution time, and their quantity increases as the number of semantic graphs grows. This results in substantial overhead and idle time from frequent kernel launches~\cite{kernel_launch_overhead}. As shown in Fig.~\ref{fig:motivation_overhead}(a), many short-execution-time CUDA kernels are used to build semantic graphs according to mini-batches. Specifically, the `compare' and `index-select' kernels, the two main kernels, are used to match edge types and select edge indices before retrieving vertex features for each semantic graph from the entire set of vertex features in the heterogeneous graph. The number of these kernels increases with the number of semantic graphs, while their execution time can be as low as 3.3 microseconds. Moreover, most kernels in this stage are typically memory-bound, exacerbating the under-utilization of compute resources, as shown in Fig.~\ref{fig:motivation_overhead}(b). This is due to their low arithmetic intensity and reliance on irregular neighbor connections~\cite{MetaNMP}.

\textbf{Inefficiency in the Neighbor Aggregation Stage.} In the neighbor aggregation stage, a set of CUDA kernels is assigned for the processing of each semantic graph. As the number of semantic graphs increases, so does the number of these kernels, compounding the overhead and idle time from frequent kernel launches. As depicted in Fig.~\ref{fig:motivation_overhead}(a), a set of short-execution-time CUDA kernels perform the aggregation of neighboring features for each vertex in each semantic graph. For example, the `scatter' and `gather' kernels are the two main kernels used to scatter the neighboring features along edges to target vertices and gather these features for each target vertex. The number of these kernels increases with the number of semantic graphs, while their execution time can be as low as 2.6 microseconds. Moreover, these kernels are typically memory-bound, exacerbating the under-utilization of compute resources. As shown in Fig.~\ref{fig:motivation_overhead}(b), most kernels used in neighbor aggregation are memory-bound due to their low arithmetic intensity and reliance on irregular neighbor connections~\cite{HiHGNN,HGNN_characterization}.

\textbf{Summary of Inefficiency Analysis.}
The combination of frequent kernel launches, short execution time, and memory bound prevents GPU from maintaining sustained high utilization, regardless of the semantic graph build and neighbor aggregation stages. This inefficiency underscores the need for optimization strategies that can reduce kernel overhead, increase execution efficiency, and ultimately improve the overall performance of GPU-based solutions for HGNNs.

\section{Design of HiFuse}

%


\subsection{Overview of HiFuse}

To address inefficiencies in previous GPU-based solutions for HGNN training, we introduce HiFuse, an enhancement for PyG aimed at accelerating mini-batch HGNN training on CPU-GPU systems. HiFuse employs a two-pronged strategy: one focusing on data reorganization and the other on workflow optimization.

From the data perspective, we propose a method for feature reorganization and merging during the neighbor aggregation stage to optimize data access and execution, thereby enhancing GPU processing efficiency. The reorganization technique adjusts the layout of vertex features by grouping features of the same vertex type together, improving memory access patterns, data locality, and overall memory efficiency. The merging technique combines smaller feature matrices into larger ones, allowing single kernels to process more data, reducing the number of kernel launches, and thus improving GPU utilization and throughput.

From the workflow perspective, we offload the edge index selection in the semantic graph build stage from GPU to CPU, thereby avoiding the overhead of numerous small CUDA kernel launches and the under-utilization of compute resources. To ensure this transition does not introduce bottlenecks, we employ two strategies: first, parallel processing on CPU handles multiple tasks concurrently using multi-threading to fully utilize CPU cores; second, an asynchronous pipeline overlaps execution stages across CPU and GPU, ensuring effective utilization of both and minimizing idle times.

\subsection{Reorganizing and Merging Features for Neighbor Aggregation}

\subsubsection{Reorganizing Features.}

In homogeneous graphs, vertex features are stored contiguously based solely on the vertex index, requiring just one dimension of organization. However, in heterogeneous graphs, vertex features involve two dimensions: vertex type and vertex index. 

\begin{figure}[!htbp]
    \centering
    \includegraphics[page=1, width=0.8\textwidth]{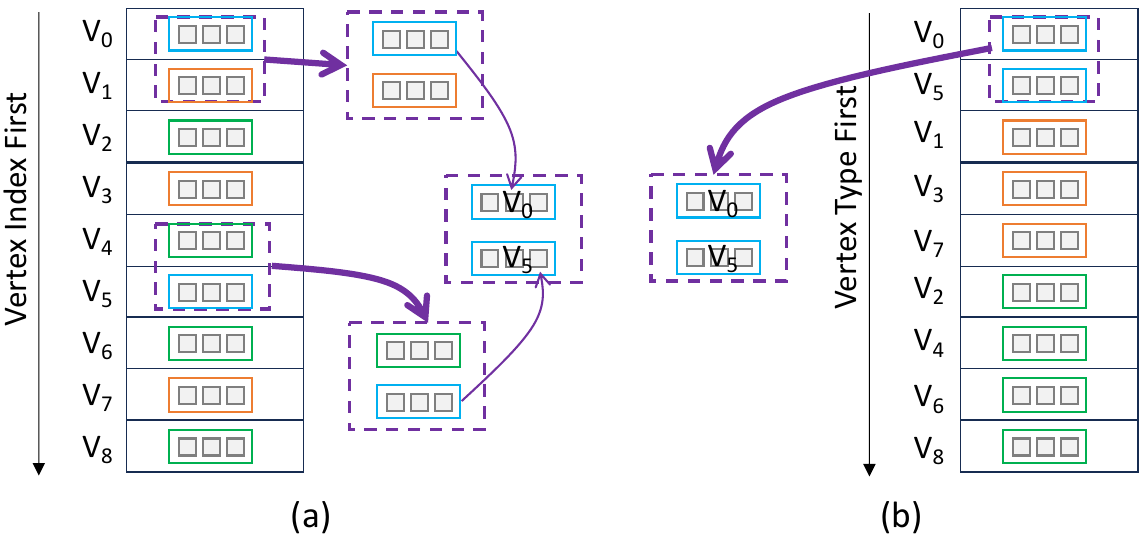}
    \vspace{-10pt}
    \caption{Data organization and accesses to vertex features in neighbor aggregation: (a) Features organized by vertex index first; (b) Features organized by vertex type first.}
    \label{fig:previous_feature_organization_and_aggregation} 
    \vspace{-10pt}
\end{figure}

Existing approaches, derived from homogeneous graph storage methods, continue to store vertex features of heterogeneous graphs primarily based on vertex index. In these approaches, vertex features are stored contiguously based on vertex indices, as illustrated in Fig.~\ref{fig:previous_feature_organization_and_aggregation}(a). This results in the features of different vertex types being stored consecutively.
This data organization leads to poor spatial and temporal locality during the neighbor aggregation. The computation for each semantic graph involves vertex features of a single vertex type~\cite{GDR_HGNN,HGNN_characterization,HiHGNN}. The interleaved storage of different types prevents the full utilization of coalesced memory access of GPU, resulting in invalid accessed data. Additionally, since cache space is occupied by features of other vertex types not required by the currently processing semantic graph, it exacerbates the replacement of needed data. Even after the feature collection, the issue of poor data locality persists.

To address this issue, the key idea is to reorganize the features based on vertex types first and then vertex indices. Features of the same type should be stored contiguously according to the order of vertices of that type, as shown in Fig.~\ref{fig:previous_feature_organization_and_aggregation}(b). This organization allows memory accesses to be coalesced, leveraging the high bandwidth of GPU device memory, thus avoiding invalid data access and frequent cache replacement.
To enhance this reorganization, we first pre-load the features into GPU device memory and offload these functions to GPU to leverage its high bandwidth. A large heterogeneous graph can be partitioned into several subgraphs, allowing the vertex features of these subgraphs to be pre-loaded into device memory. We implement a CUDA kernel to reorganize and retrieve features according to the mini-batch for neighbor aggregation.

Our proposed data reorganization improves memory access patterns by enhancing data locality, ensuring that computations for each semantic graph can access and efficiently utilize the relevant vertex features.

\textbf{Merging Features.} Previous efforts treat the computation of multiple semantic graphs as independent homogeneous GNN computations. While this method completes the required tasks, it introduces numerous CUDA kernels, significantly increasing kernel launch overhead.

Typically, kernel fusion is used to reduce the number of kernels by merging the functionality of different kernels, thereby reducing intermediate data access and streamlining operations. In our scenario, however, the kernels are identical and lack dataflow dependencies. Therefore, a different approach is required to address this situation effectively. We tackle this problem from a data perspective by implementing a data merging technique. This involves combining smaller feature matrices from different semantic graphs into larger ones, which allows us to modify the input size of the kernels. By increasing the input data size of each kernel, we can significantly reduce the total number of kernels required. This approach not only reduces the overhead associated with frequent kernel launches but also improves GPU utilization and overall processing throughput.

\begin{figure}[!htbp]
    \vspace{-10pt}
    \centering
    \includegraphics[page=1, width=0.785\textwidth]{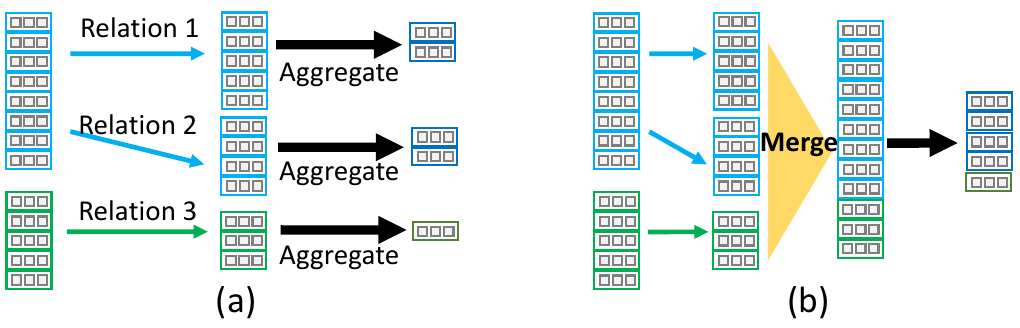}
    \vspace{-10pt}
    \caption{Neighbor aggregation (a) without feature merging and (b) with feature merging.}
    \label{fig:HiFuse_feature_merging_and_aggregation} 
    \vspace{-10pt}
\end{figure}

Fig.~\ref{fig:HiFuse_feature_merging_and_aggregation}(a) and (b) respectively illustrate the differences in neighbor aggregation without and with feature merging. Without feature merging, each semantic graph performs its own aggregation computation separately, producing the aggregated vertex features individually. With feature merging, the features of source vertices from all semantic graphs are extracted and combined into a single tensor for a unified aggregation computation, yielding the aggregated vertex features in a single step. This significantly reduces the number of CUDA kernels required and minimizes the time overhead associated with neighbor aggregation.

\begin{algorithm}[!t]
    \SetAlgoLined
    \caption{\textbf{Neighbor Aggregation with Reorganization and Merging}}  \label{alg:aggregation_optimization}
    \textbf{input:} Vertex Features $x$, Source Vertex Type $SrcType$,  \\
    Source Vertex Index $SrcIndex$, Destination Vertex Index $DstIndex$, \\
    Number of Semantic Graphs $NumSemanticGraph$ \\
    \textbf{output:} Aggregated Vertex Features $AggrResults$ \\
    \For{i $\gets$ 0 \textbf{to} $NumSemanticGraph - 1$}{
       Features $\gets$ IndexSelect($x$[$SrcType$[i]], $SrcIndex$[i]) \\
       FeaturesArray.Append(Features) \\
       DstIndexArray.Append($DstIndex$[i]) 
    }
    FeatureCat $\gets$ Concat(FeaturesArray) \\
    DstIndexCat $\gets$ Concat(DstIndexArray) \\
    $AggrResults$ $\gets$ Aggregate(FeatureCat, DstIndexCat) \\
    \textbf{return} $AggrResults$
\end{algorithm}

The implementation of neighbor aggregation using feature reorganization and merging is detailed in Algorithm \ref{alg:aggregation_optimization}. The data required for neighbor aggregation includes vertex features $x$, the types of source vertices $SrcType$, the source vertex indices $SrcIndex$, and the destination vertex indices $DstIndex$ of all semantic graphs. Unlike computing the neighbor aggregation for each semantic graph separately, this process temporarily stores the collected source vertex features in a container after each semantic graph completes its feature collection. This approach exploits the data locality provided by the feature reorganization. Similarly, the destination vertex indices for each semantic graph are also temporarily stored.
Once feature collection for all semantic graphs is complete, the discrete source vertex features in the container are merged into a single tensor, and the destination vertex indices are merged similarly. These merged tensors are then used as inputs for the aggregate function, which computes the aggregated vertex features for the destination vertices of each semantic graph.

The feature reorganization and merging method not only enhances data locality and memory access efficiency but also reduces the number of CUDA kernel launches, thereby improving GPU utilization and overall processing throughput. It is particularly effective for heterogeneous graph datasets with numerous relations (semantic graphs). By handling larger chunks of data with fewer, more efficient kernels, we minimize idle times, reduce computational overheads, and maximize the use of GPU resources.

\subsection{Offloading and Parallelizing Edge Index Selection for Semantic Graph Build}

\textbf{Observations.} 
After sampling, each mini-batch mainly includes the edge indices of the sampled subgraph. Each edge index consists of the IDs of the source vertex and target vertex of an edge. These edge indices are stored in a $2 \times N$ tensor in coordinate format, representing the graph topology for all relations. Before processing each semantic graph, the edge indices of the same edge type must be selected for each semantic graph, a process known as edge index selection in the semantic graph build stage. This involves edge type matching and index selection, carried out by the compare and index-select operations.

Observation \blackcircled{1}: The kernels used for compare and index-select operations exhibit high control intensity and a high proportion of integer instructions because they primarily involve integer comparison and addressing operations. Additionally, these kernels have low computational demands and short execution times, resulting in underutilization of GPU's processing capabilities.

Observation \blackcircled{2}: During each mini-batch training process, aside from data sampling on CPU, all other computations occur on GPU. Consequently, CPU remains mostly idle, leading to an imbalance in the utilization of CPU and GPU. As shown in Table~\ref{table:observation}, the CPU utilization is generally lower than that of GPU. The ratio of CPU time to GPU time is as low as 0.04.

\begin{table}[!htbp]
    \centering
    \caption{Execution time of CPU and GPU in one epoch of training.}
    \label{table:observation} 
    \renewcommand\arraystretch{1.2}
    \setlength{\tabcolsep}{8pt}
    \begin{tabular}{ccccc}
    \toprule
    & \multicolumn{2}{c}{\textbf{RGCN-AM}} & \multicolumn{2}{c}{\textbf{RGAT-AM}} \\ \midrule
    \multicolumn{1}{c}{} & \multicolumn{1}{c}{Time (ms)} & Ratio                 & \multicolumn{1}{c}{Time (ms)} & Ratio                 \\ 
    \textbf{CPU}                    & \multicolumn{1}{c}{131}        & \multirow{2}{*}{0.13} & \multicolumn{1}{c}{133}        & \multirow{2}{*}{0.04}\\ 
    \textbf{CPU} & \multicolumn{1}{c}{998}  &  & \multicolumn{1}{c}{3177}  &  \\ \bottomrule
    \end{tabular}
\end{table}

\textbf{Offloading.} 
Based on our observations, we offload the kernels for edge index selection from GPU to CPU, driven by several key considerations as follows.

\squishlist
 \item Enhanced Resource Utilization: CPU is underutilized, with most workloads assigned to GPU, leaving it mostly idle. By offloading control-intensive tasks to CPU, we can balance the workload between CPU and GPU, making better use of available resources.
   
 \item Reduced Overhead: Kernels with short execution time do not significantly benefit from GPU acceleration and can incur overhead from frequent kernel launches. Executing these tasks on control-friendly CPU can eliminate this overhead, leading to more efficient processing.

\squishend

\begin{algorithm}[!t]
    \SetAlgoLined
    \caption{\textbf{Edge Index Selection}} \label{alg:edge_index_selection_optimization}
    \textbf{input:} Edge Index $EdgeIndex$, Edge ID $EdgeID$, Edge Type $EdgeType$,\\ Number of GNN Layer $NumLayer$, Number of Semantic Graphs $NumSemanticGraph$ \\
    \textbf{output:} Edge Index Array $EdgeIndexArray$ \\
    \For{i $\gets$ 0 \textbf{to} $NumLayer - 1$}{
       $EdgeTypeLayer$ $\gets$ IndexSelect($EdgeType$, $EdgeID$[i]) \\
       \For{j $\gets$ 0 \textbf{to} $NumSemanticGraph - 1$}{
            mask $\gets$ compare(j, EdgeTypeLayer) \\
            TempEdgeIndex $\gets$ IndexSelect($EdgeIndex$[i], mask) \\
            $EdgeIndexArray$[i].Append(TempEdgeIndex) 
       }
    }
    \textbf{return} $EdgeIndexArray$
\end{algorithm}

Since each semantic graph's computation in each layer requires edge indices, it is most efficient to perform edge index selection after sampling but before model computation. The edge indices for all relations are temporarily stored, allowing direct access during computation without further indexing. The specific implementation is detailed in Algorithm \ref{alg:edge_index_selection_optimization}. Edge indices (EdgeIndex) and edge identifiers (EdgeID) are obtained from the mini-batch, representing the graph topology and the identifiers of edges within the sampled subgraph, respectively. Before the indexing computation for each layer, the EdgeID is used to index all edge types within the sampled subgraph for that layer, resulting in a tensor of edge types. During the indexing operation for each relation, edge type matching is first performed using the compare operation to select indices of the same edge type from the edge type tensor. Subsequently, the index-select operation is used to select the edge index tensor for the current relation based on these indices. Finally, the indices edge tensor for each relation is stored, completing the edge index selection.

By isolating edge index selection from each semantic graph's computation, the forward pass retains only those operations demanding significant GPU resources. This approach improves the continuity and compactness of CUDA kernel scheduling, thus enhancing overall GPU utilization and computational efficiency during model computation.

\textbf{Parallelizing.} Offloading tasks to CPU is a relatively straightforward idea, but maximizing the CPU's parallel processing capabilities is crucial to ensure timely data transfer to GPU. To achieve this, it is necessary to leverage idle CPU cores to parallelize the edge index selection, compensating for the CPU's slower computation speed and improving overall efficiency.

The edge type matching and index selection for each relation are independent, making them ideal for parallel processing. Thus, we utilize the multi-threading capabilities of CPU to parallelize this process. This is implemented in PyTorch's CPP backend environment, LibTorch, using the OpenMP multi-threading parallel programming framework. The specific implementation follows a similar approach to Algorithm \ref{alg:edge_index_selection_optimization}.
After obtaining the EdgeIndex and EdgeID from mini-batch graph data sampling, we use OpenMP precompiled directives to enable multi-threaded parallel acceleration in the critical code segments. By employing appropriate OpenMP directives based on the CPU's multi-core resources, we enhance the performance of the multi-threaded program, ensuring efficient edge index selection and timely data transfer to GPU.

\begin{figure}[!t]
    \centering
    \includegraphics[page=1, width=0.8\textwidth]{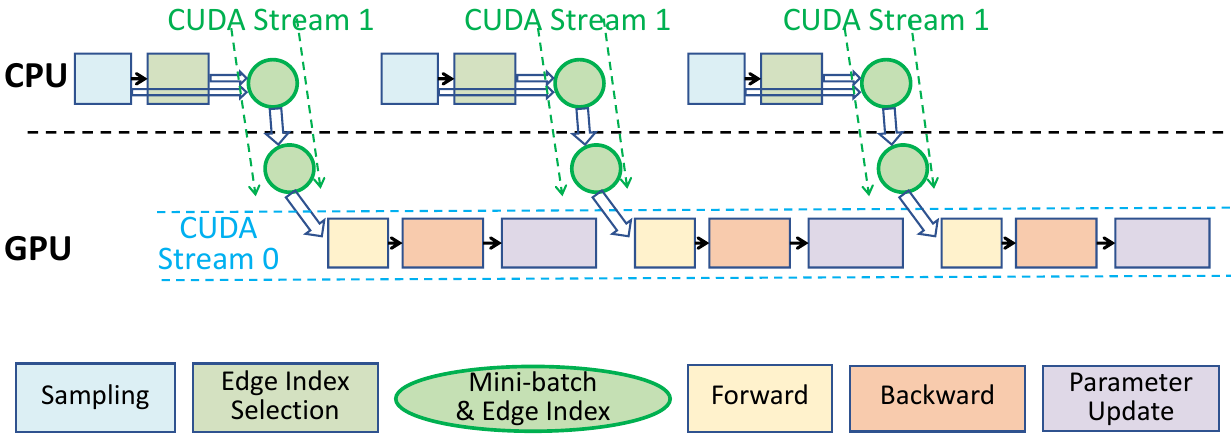}
    \vspace{-5pt}
    \caption{Pipelining execution stages across CPU and GPU.}
    \label{fig:asyn_pipeline_stages} 
    \vspace{-20pt}
\end{figure}

\textbf{Pipelining.} By combining parallel processing with an asynchronous pipeline, we can enhance the utilization of both CPU and GPU. This approach minimizes idle times, reduces latency, and maintains a steady flow of data, leading to significant performance improvements in complex computations.

Based on the above parallelization strategy, we propose an asynchronous execution method for CPU and GPU tasks to address the issue where CPU and GPU tasks in each mini-batch training session are executed sequentially, with GPU tasks taking significantly longer than CPU tasks. The optimized asynchronous computation flow is depicted in Fig.~\ref{fig:asyn_pipeline_stages}. Initially, the mini-batch sampling and edge index selection tasks are scheduled by a CPU thread. The processed mini-batch and edge index array are used for the model computation of the mini-batch. Another CPU thread transfers the mini-batch and edge index array from CPU to GPU. Data transfer is managed by a dedicated CUDA stream, which operates independently of the default CUDA stream used for model computation. Subsequently, the model computation begins on GPU. The computation tasks on GPU proceed asynchronously, including forward, backward, and parameter update for the current training mini-batch, managed by the default CUDA stream. After GPU computation tasks are completed, the CPU thread gathers and returns the results. This cycle continues until all mini-batch computations are finished.

In this CPU and GPU asynchronous mini-batch training pipeline, CPU and GPU computations can be executed in parallel. Each mini-batch's sampling and edge index selection tasks are completed before the model computation tasks begin, ensuring seamless transitions between GPU computation tasks and avoiding GPU idle states. This approach also efficiently utilizes idle CPU computing resources. Overall, this method improves the efficiency of each mini-batch training, accelerating the entire training process of the HGNN model.

\section{Results}

\subsection{Evaluation Setup} \label{sec:Methodology}

For our baseline comparison, we select PyG~\cite{PyTorch_Geometric} for the following reasons. PyG is a widely used framework for training HGNNs and undergoes extensive performance optimizations in an open-source environment, resulting in significant kernel optimizations and recent performance improvements. Therefore, it represents the current state-of-the-art GPU-based solution for HGNN training.

Both HiFuse and PyG are executed on a Linux server equipped with an Intel Xeon Silver 4208 CPU and an NVIDIA T4 GPU, utilizing single precision floats. We utilize the latest versions of PyG (2.5.3) and ensure consistent runtime environments (CUDA 12.1 with driver version 530.30.02) across all solutions.

\begin{table}[!b]
   \vspace{-15pt}
	\caption{Benchmark datasets.}\label{table:dataset}
	\centering
	\renewcommand\arraystretch{1.2}
        \setlength{\tabcolsep}{8pt}
    \resizebox{1\textwidth}{!}{
		\begin{tabular}{*6{c}}
	     	\toprule   
			\textbf{Dataset}          & \textbf{\#Nodes}   &  \textbf{\#Edges}  & \textbf{\#Node Types}   &  \textbf{\#Edge Relations} \\  \midrule  
  			 aifb (AF)               & 7,262            & 48,810           & 7            &  104 \\ 
  			 bgs  (BG)               & 94,806           & 672,884          & 27           &  122 \\ 
                  mutag  (MT)             & 27,163           & 148,100          & 5            &  50 \\ 
                  am (AM)                 & 1,885,136        & 5,668,682        & 7            &  108 \\ 
             \bottomrule
		\end{tabular}
	}
    \vspace{-15pt}
\end{table}

For evaluation, we employ two well-known HGNN models: RGCN~\cite{R-GCN}, representing an HGNN model with a simple architecture, and RGAT~\cite{R-GAT}, representing an HGNN model with a complex architecture. Additionally, we utilize four popular datasets (aifb, mutag, bgs, and am)~\cite{R-GCN}, as shown in Table~\ref{table:dataset}.

To assess performance, we utilize NVIDIA Nsight System and Nsight Compute. Each model, along with its respective dataset, is executed ten times, and outlier results are discarded for reliability. These tools offer detailed insights into the performance metrics of our models, facilitating accurate analysis and comparison of efficiency and optimization benefits.

\subsection{Comparison with State-of-the-Art HGNN Training Framework}

\begin{wrapfigure}{l}{0.5\textwidth}
    \centering
    \includegraphics[page=1, width=0.5\textwidth]{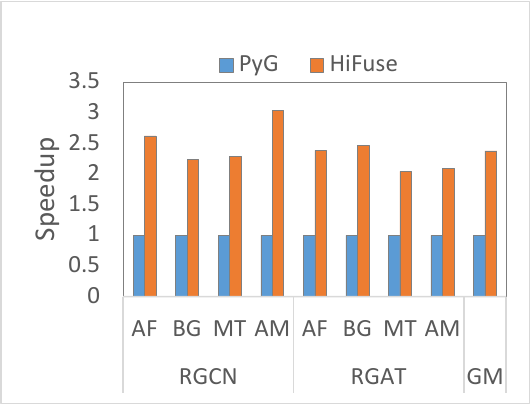}
    \vspace{-20pt}
    \caption{Speedup to state-of-the-art GPU-based solution (PyG) across different datasets and models.} \label{fig:overall_speedup} 
    \vspace{-15pt}
\end{wrapfigure}

\textbf{Speedup.} To demonstrate the effectiveness of our optimizations, we compare HiFuse with the state-of-the-art framework, PyG. Both HiFuse and PyG pre-load vertex features into GPU device memory. For large heterogeneous graphs, they can be partitioned into several subgraphs, enabling the vertex features of these subgraphs to be pre-loaded into device memory. In the comparison, GM represents the geometric mean. As shown in Fig.~\ref{fig:overall_speedup}, HiFuse achieves an average performance improvement of 2.38$\times$ compared to PyG, with a maximum improvement of up to 3.04$\times$. 
The improvement primarily comes from the following reasons: 
1) The number of GPU kernels is significantly reduced, resulting in a substantial decrease in kernel launch time and idle time.
2) The pipeline and parallel execution between CPU and GPU ensures that CPU promptly provides input data to GPU.
3) The workload is balanced between CPU and GPU to some extent, enhancing overall efficiency.

\begin{wrapfigure}{l}{0.5\textwidth}
    \vspace{-20pt}
    \centering
    \includegraphics[page=1, width=0.5\textwidth]{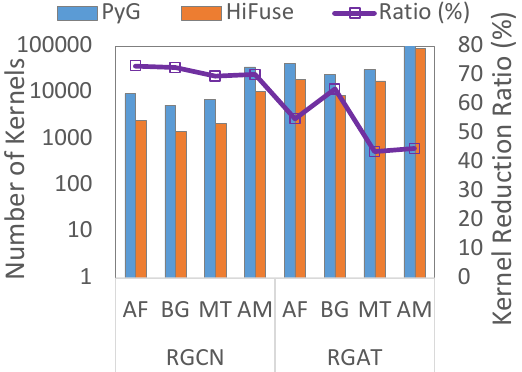}
    \vspace{-20pt}
    \caption{Number of kernels and its reduction ratio in one epoch using HiFuse compared to PyG on different datasets and models.}
    \label{fig:kernel_reduction_ratio} 
    \vspace{-25pt}
\end{wrapfigure}

\textbf{Number of Kernels.} To demonstrate the effect of reducing the number of kernels, we collect kernel counts for the training of one epoch. As shown in Fig.~\ref{fig:kernel_reduction_ratio}, HiFuse achieves a kernel reduction ratio of 43.6\% to 73.2\% compared to PyG across different datasets and models. This significant reduction is attributed to our offloading and merging optimizations. The reduction ratio in the RGAT model is smaller than in the RGCN model due to the additional kernels required for attention calculation.

\subsection{Optimization Effect Analysis}

\begin{figure}[!htbp]
    \centering
    \includegraphics[page=1, width=0.95\textwidth]{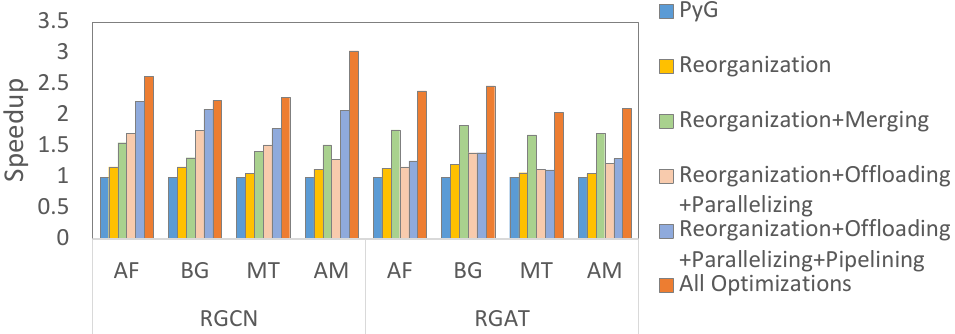}
    \vspace{-10pt}
    \caption{Speedup to baseline (PyG) on different optimization configurations across different datasets and models.}
    \label{fig:speedup_breakdown} 
    \vspace{-20pt}
\end{figure}

\textbf{Speedup.} To evaluate the effect of each optimization on performance, we conduct an ablation study. The results, shown in Fig.~\ref{fig:speedup_breakdown}, demonstrate that, compared to the baseline, using only the reorganization optimization results in a performance improvement of up to 1.17$\times$. This is due to improved data locality. Using both the reorganization and merging optimizations yields a performance increase of up to 1.83$\times$, benefiting from reduced kernel launch overhead and further improvements in data locality.
Applying the reorganization, offloading, and parallelization optimizations results in a speedup of up to 1.7$\times$ due to reduced kernel launch time, improved data locality, and parallel execution. Additionally, incorporating the pipelining optimization achieves a speedup of up to 2.2$\times$.

An interesting observation is that the RGAT model benefits more from the merging optimization, while the RGCN model gains more from the offloading, parallelizing, and pipelining optimizations. This is because the RGAT model has a more computation-intensive model computation phase, making the GPU's execution time more dominant. Therefore, optimizations that enhance the neighbor aggregation stage on GPU have a greater effect on reducing the execution time of the RGAT model. On the other hand, the RGCN model has a less computation-intensive model computation phase, meaning the GPU's execution time is not as dominant. Consequently, optimizations targeting the CPU-executed phases are more effective for the RGCN model.

\begin{wrapfigure}{l}{0.5\textwidth}
    \vspace{-20pt}
    \centering
    \includegraphics[page=1, width=0.5\textwidth]{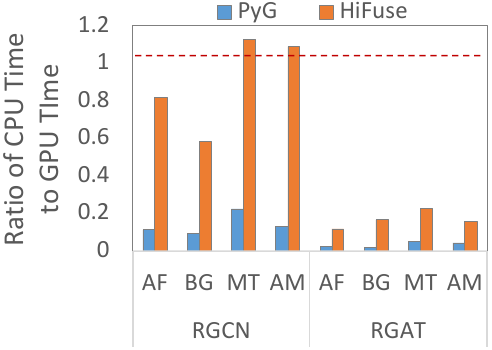}
    \vspace{-20pt}
    \caption{Ratio of CPU's execution time to GPU's execution time.}
    \label{fig:ratio_cpu_gpu_time} 
    \vspace{-15pt}
\end{wrapfigure}

\textbf{Ratio of CPU Time to GPU Time.}
To gain deeper insights into our optimizations, we analyze the ratio of CPU execution time to GPU execution time. The closer this ratio is to one, the more balanced the utilization of CPU and GPU, indicating more effective pipelined execution. As shown in Fig.~\ref{fig:ratio_cpu_gpu_time}, HiFuse achieves a ratio closer to one compared to PyG across different datasets and models. This indicates that our optimizations balance CPU and GPU utilization more effectively, resulting in more efficient pipelined execution. HiFuse achieves this by offloading edge index selection from GPU to CPU, parallelizing this process, and effectively bridging the dataflow between CPU and GPU with pipeline optimizations. Additionally, we observe that the impact differs between the RGAT and RGCN models. The RGAT model, which has a more computation-intensive model computation phase, benefits differently from these optimizations compared to the RGCN model.

\begin{wrapfigure}{l}{0.5\textwidth}
    \centering
    \includegraphics[page=1, width=0.5\textwidth]{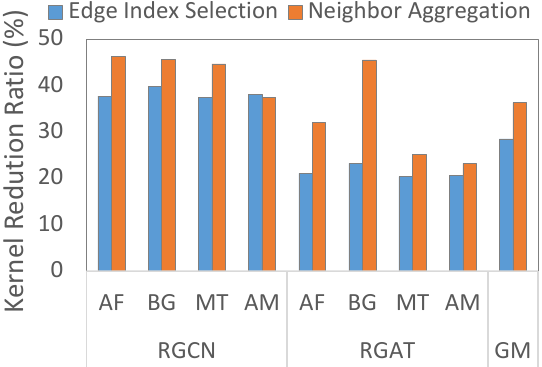}
    \vspace{-20pt}
    \caption{Reduction ratio of number of kernels in edge index selection and neighbor aggregation during forward pass when using HiFuse compared to PyG across different datasets and models.}
    \label{fig:kernel_reduction_ratio_breakdown} 
    \vspace{-10pt}
\end{wrapfigure}

\textbf{Number of Kernel Reduction.} To illustrate the effectiveness of the offloading and merging optimizations, we delve into the edge index selection and neighbor aggregation stages. As shown in Fig.~\ref{fig:kernel_reduction_ratio_breakdown}, offloading edge index selection from GPU to CPU can reduce the number of kernels by up to 39.8\%. Additionally, merging features to enable a single kernel for neighbor aggregation can reduce the number of kernels by up to 46.3\%.

\textbf{Compute and Memory Throughput.} To demonstrate the optimization achieved through feature reorganization and merging, we investigate GPU compute and memory throughput. Compute throughput measures the efficiency of GPU computational resource utilization, while memory throughput indicates the efficiency of data transfer between the GPU's memory and its computational units. As shown in Table~\ref{tab:compute_memory_throughput}, compared to PyG, the compute and memory throughput of the main kernel (`scatter') used in the neighbor aggregation stage are improved by up to 136$\times$ and 14$\times$, respectively. This improvement stems from feature reorganization and merging the inputs of multiple kernels into a single GPU kernel, offering several benefits:
1) Larger kernel sizes can better leverage the GPU's parallel processing capabilities.
2) Fewer kernels significantly reduce kernel launch time overhead.
3) Enhanced data locality improves coalesced memory access and reduces memory access latency in the GPU pipeline, enabling faster data supply.

\begin{table}[!b]
\vspace{-15pt}
\centering
	\renewcommand\arraystretch{1.2}
        \setlength{\tabcolsep}{5pt}
\caption{Compute throughput and memory throughput of `scatter' kernel.} \label{tab:compute_memory_throughput}
\begin{tabular}{ccccccc}
\toprule
 & \multicolumn{2}{|c|}{\textbf{PyG}}           & \multicolumn{2}{c|}{\textbf{HiFuse}}          & \multicolumn{2}{c}{\textbf{Improvement Ratio}} \\ \midrule
 & \multicolumn{1}{c}{Compute} & Memory & \multicolumn{1}{c}{Compute} & Memory & \multicolumn{1}{c}{Compute} & Memory \\ 
\textbf{RGCN-AM} & \multicolumn{1}{c}{0.15\%} & 2.04\% & \multicolumn{1}{c}{14.42\%} & 23.42\% & \multicolumn{1}{c}{95}     & 11    \\ 
\textbf{RGAT-AM} & \multicolumn{1}{c}{0.14\%} & 2.16\% & \multicolumn{1}{c}{19.41\%} & 29.95\% & \multicolumn{1}{c}{136}    & 14    \\ \bottomrule
\end{tabular}
\vspace{-30pt}
\end{table}

\section{Related Work} \label{sec:related_work}

Currently, little work is focused on accelerating HGNNs using CPUs and/or GPUs, with main efforts concentrated in PyG~\cite{PyTorch_Geometric}, DGL~\cite{DistDGLv2}, and HGL~\cite{HGL}. PyG~\cite{PyTorch_Geometric} provides a unified framework for handling multiple vertex and edge types, optimized sampling methods to reduce memory usage, and type-specific operations for customized transformations. By leveraging PyTorch's tensor operations and GPU acceleration, PyG~\cite{PyTorch_Geometric} ensures efficient computations, making it a powerful tool for processing complex, multi-relational data in heterogeneous graphs. DGL~\cite{DistDGLv2} optimizes HGNN training by reducing data movement between distributed CPU memory and GPUs, balancing mini-batch loads, and maximizing hardware utilization. HGL~\cite{HGL} introduces an intermediate representation called HIR, translates user-constructed GNN models into HIR, and applies optimizations such as graph stitching, kernel fusion, and operator bundling to address memory fragmentation and leverage parallelism across semantic graphs.

Similarly, there is little work focused on accelerating HGNNs using domain-specific hardware accelerators, with main efforts in MetaNMP~\cite{MetaNMP}, HiHGNN~\cite{HiHGNN}, GDR-HGNN~\cite{GDR_HGNN}, and ADE-HGNN~\cite{ADE-HGNN}. MetaNMP~\cite{MetaNMP} proposes a DIMM-based near-memory processing HGNN accelerator, which reduces memory footprint and improves performance by generating metapath instances on-the-fly to avoid intermediate storage. HiHGNN~\cite{HiHGNN} proposes a high-performance HGNN hardware accelerator using bound-aware stage fusion, independency-aware parallel execution, and similarity-aware execution scheduling to enhance performance. GDR-HGNN~\cite{GDR_HGNN} introduces a hardware frontend that dynamically restructures heterogeneous graphs to enhance data locality and address buffer thrashing during HGNN acceleration. ADE-HGNN~\cite{ADE-HGNN} designs an HGNN accelerator to exploit attention disparity through a runtime pruning method based on min-heap, aimed at discarding unimportant vertices.

Unlike previous work, our work accelerates HGNN training by reducing the number of short-execution-time and memory-bound CUDA kernels on GPU. Importantly, we achieve this without kernel fusion; instead, we strategically offload CUDA kernels from GPU to CPU and merge input data from multiple identical kernels to enable the use of a single kernel. Additionally, our method is orthogonal and complementary to techniques such as kernel fusion.

\section{Conclusion}

In this work, we accelerate mini-batch HGNN training by introducing HiFuse, an enhancement for PyG. HiFuse reorganizes and merges vertex features, and offloads CUDA kernels with low computational demands from GPU to CPU. This approach significantly reduces the number of CUDA kernels and improves GPU utilization. Additionally, HiFuse leverages parallelization and asynchronous pipelines to ensure seamless execution between CPU and GPU, reducing idle times and improving resource usage for both processors. Through extensive experiments with various HGNN models and datasets, we demonstrate that HiFuse achieves substantial performance gains compared to PyG.

\textbf{Acknowledgments.} This work was supported in part by National Key Research and Development Program under Grant 2022YFB4501400, in part by the National Natural Science Foundation of China under Grant 62202451, in part by CAS Project for Young Scientists in Basic Research under Grant YSBR-029, and in part by CAS Project for Youth Innovation Promotion Association.

\bibliographystyle{IEEEtran}
\bibliography{refs}

\begin{thebibliography}{10}
\providecommand{\url}[1]{#1}
\csname url@samestyle\endcsname
\providecommand{\newblock}{\relax}
\providecommand{\bibinfo}[2]{#2}
\providecommand{\BIBentrySTDinterwordspacing}{\spaceskip=0pt\relax}
\providecommand{\BIBentryALTinterwordstretchfactor}{4}
\providecommand{\BIBentryALTinterwordspacing}{\spaceskip=\fontdimen2\font plus
\BIBentryALTinterwordstretchfactor\fontdimen3\font minus \fontdimen4\font\relax}
\providecommand{\BIBforeignlanguage}[2]{{%
\expandafter\ifx\csname l@#1\endcsname\relax
\typeout{** WARNING: IEEEtran.bst: No hyphenation pattern has been}%
\typeout{** loaded for the language `#1'. Using the pattern for}%
\typeout{** the default language instead.}%
\else
\language=\csname l@#1\endcsname
\fi
#2}}
\providecommand{\BIBdecl}{\relax}
\BIBdecl

\bibitem{oh2018knowledge}
B.~Oh, S.~Seo, and K.-H. Lee, ``Knowledge graph completion by context-aware convolutional learning with multi-hop neighborhoods,'' in \emph{Proceedings of the 27th ACM International Conference on Information and Knowledge Management}, 2018.

\bibitem{zhang2019iteratively}
W.~Zhang, B.~Paudel, L.~Wang, J.~Chen, H.~Zhu, W.~Zhang, A.~Bernstein, and H.~Chen, ``Iteratively learning embeddings and rules for knowledge graph reasoning,'' in \emph{The world wide web conference}, 2019.

\bibitem{chen2017task}
T.~Chen and Y.~Sun, ``Task-guided and path-augmented heterogeneous network embedding for author identification,'' in \emph{Proceedings of the tenth ACM international conference on web search and data mining}, 2017, pp. 295--304.

\bibitem{yasunaga2019scisummnet}
M.~Yasunaga, J.~Kasai, R.~Zhang, A.~R. Fabbri, I.~Li, D.~Friedman, and D.~R. Radev, ``Scisummnet: A large annotated corpus and content-impact models for scientific paper summarization with citation networks,'' in \emph{Proceedings of the AAAI conference on artificial intelligence}, 2019.

\bibitem{zheng2020clustering}
Y.~Zheng, R.~Hu, S.-f. Fung, C.~Yu, G.~Long, T.~Guo, and S.~Pan, ``Clustering social audiences in business information networks,'' \emph{Pattern Recognition}, 2020.

\bibitem{hgnn_survey_tangjie}
Y.~Dong, Z.~Hu, K.~Wang, Y.~Sun, and J.~Tang, ``Heterogeneous network representation learning.'' in \emph{IJCAI}, vol.~20, 2020, pp. 4861--4867.

\bibitem{hgnn_survey_hanjiawei}
C.~Yang, Y.~Xiao, Y.~Zhang, Y.~Sun, and J.~Han, ``Heterogeneous network representation learning: A unified framework with survey and benchmark,'' \emph{IEEE Transactions on Knowledge and Data Engineering}, 2020.

\bibitem{hgnn_survey_shichuan}
X.~Wang, D.~Bo, C.~Shi, S.~Fan, Y.~Ye, and P.~S. Yu, ``A survey on heterogeneous graph embedding: methods, techniques, applications and sources,'' \emph{arXiv preprint arXiv:2011.14867}, 2020.

\bibitem{hgnn_survey_shiruipan}
X.~Zheng, Y.~Liu, S.~Pan, M.~Zhang, D.~Jin, and P.~S. Yu, ``Graph neural networks for graphs with heterophily: A survey,'' \emph{arXiv preprint arXiv:2202.07082}, 2022.

\bibitem{li2022disentangled}
A.~Li, Z.~Cheng, F.~Liu, Z.~Gao, W.~Guan, and Y.~Peng, ``Disentangled graph neural networks for session-based recommendation,'' \emph{IEEE Transactions on Knowledge and Data Engineering}, 2022.

\bibitem{fan2019metapath}
S.~Fan, J.~Zhu, X.~Han, C.~Shi, L.~Hu, B.~Ma, and Y.~Li, ``Metapath-guided heterogeneous graph neural network for intent recommendation,'' in \emph{Proceedings of the 25th ACM SIGKDD international conference on knowledge discovery \& data mining}, 2019.

\bibitem{luo2021imas}
F.~Luo, Y.~Zhang, and X.~Wang, ``Imas++ an intelligent medical analysis system enhanced with deep graph neural networks,'' in \emph{Proceedings of the 30th ACM International Conference on Information \& Knowledge Management}, 2021.

\bibitem{CompGCN}
S.~Vashishth, S.~Sanyal, V.~Nitin, and P.~Talukdar, ``Composition-based multi-relational graph convolutional networks,'' in \emph{International Conference on Learning Representations}, 2020.

\bibitem{M2GNN}
S.~Wang, X.~Wei, C.~N. Nogueira~dos Santos, Z.~Wang, R.~Nallapati, A.~Arnold, B.~Xiang, P.~S. Yu, and I.~F. Cruz, ``Mixed-curvature multi-relational graph neural network for knowledge graph completion,'' in \emph{Proceedings of the Web Conference 2021}, 2021.

\bibitem{liu2018heterogeneous}
Z.~Liu, C.~Chen, X.~Yang, J.~Zhou, X.~Li, and L.~Song, ``Heterogeneous graph neural networks for malicious account detection,'' in \emph{Proceedings of the 27th ACM international conference on information and knowledge management}, 2018.

\bibitem{mao2020item}
K.~Mao, X.~Xiao, J.~Zhu, B.~Lu, R.~Tang, and X.~He, ``Item tagging for information retrieval: a tripartite graph neural network based approach,'' in \emph{Proceedings of the 43rd International ACM SIGIR Conference on Research and Development in Information Retrieval}, 2020, pp. 2327--2336.

\bibitem{HGNN_characterization}
M.~Yan, M.~Zou, X.~Yang, W.~Li, X.~Ye, D.~Fan, and Y.~Xie, ``Characterizing and understanding {HGNNs} on {GPUs},'' \emph{IEEE Computer Architecture Letters}, 2022.

\bibitem{HiHGNN}
R.~Xue, D.~Han, M.~Yan, M.~Zou, X.~Yang, D.~Wang, W.~Li, Z.~Tang, J.~Kim, X.~Ye, and D.~Fan, ``{HiHGNN}: Accelerating {HGNNs} through parallelism and data reusability exploitation,'' \emph{IEEE Transactions on Parallel and Distributed Systems}, 2024.

\bibitem{SeHGNN}
X.~Yang, M.~Yan, S.~Pan, X.~Ye, and D.~Fan, ``Simple and efficient heterogeneous graph neural network,'' in \emph{Proceedings of the AAAI Conference on Artificial Intelligence}, vol.~37, no.~9, 2023, pp. 10\,816--10\,824.

\bibitem{HGL}
Y.~Gui, Y.~Wu, H.~Yang, T.~Jin, B.~Li, Q.~Zhou, J.~Cheng, and F.~Yu, ``{HGL}: accelerating heterogeneous {GNN} training with holistic representation and optimization,'' in \emph{SC22: International Conference for High Performance Computing, Networking, Storage and Analysis}.\hskip 1em plus 0.5em minus 0.4em\relax IEEE, 2022, pp. 1--15.

\bibitem{PyTorch_Geometric}
M.~Fey and J.~E. Lenssen, ``Fast graph representation learning with {PyTorch Geometric},'' in \emph{ICLR Workshop on Representation Learning on Graphs and Manifolds}, 2019.

\bibitem{DistDGLv2}
D.~Zheng, X.~Song, C.~Yang, D.~LaSalle, and G.~Karypis, ``Distributed hybrid cpu and gpu training for graph neural networks on billion-scale heterogeneous graphs,'' in \emph{Proceedings of the 28th ACM SIGKDD Conference on Knowledge Discovery and Data Mining}, 2022, pp. 4582--4591.

\bibitem{kernel_launch_overhead}
L.~Zhang, M.~Wahib, and S.~Matsuoka, ``Understanding the overheads of launching cuda kernels,'' \emph{ICPP19}, pp. 5--8, 2019.

\bibitem{HyGCN}
M.~Yan, L.~Deng, X.~Hu, L.~Liang, Y.~Feng, X.~Ye, Z.~Zhang, D.~Fan, and Y.~Xie, ``{HyGCN}: A {GCN} accelerator with hybrid architecture,'' in \emph{2020 IEEE International Symposium on High Performance Computer Architecture (HPCA)}, 2020.

\bibitem{GCN_Charaterization_CAL}
M.~Yan, Z.~Chen, L.~Deng, X.~Ye, Z.~Zhang, D.~Fan, and Y.~Xie, ``Characterizing and understanding {GCNs} on {GPU},'' \emph{IEEE Computer Architecture Letters}, 2020.

\bibitem{MetaNMP}
D.~Chen, H.~He, H.~Jin, L.~Zheng, Y.~Huang, X.~Shen, and X.~Liao, ``{MetaNMP}: leveraging cartesian-like product to accelerate {HGNNs} with near-memory processing,'' in \emph{Proceedings of the 50th Annual International Symposium on Computer Architecture}, 2023.

\bibitem{HG_survey}
C.~Shi, Y.~Li, J.~Zhang, Y.~Sun, and S.~Y. Philip, ``A survey of heterogeneous information network analysis,'' \emph{IEEE Transactions on Knowledge and Data Engineering}, vol.~29, no.~1, pp. 17--37, 2016.

\bibitem{GNN_DT_Survey}
H.~Lin, M.~Yan, X.~Ye, D.~Fan, S.~Pan, W.~Chen, and Y.~Xie, ``A comprehensive survey on distributed training of graph neural networks,'' \emph{Proceedings of the IEEE}, 2023.

\bibitem{GDR_HGNN}
R.~Xue, M.~Yan, D.~Han, Y.~Teng, Z.~Tang, X.~Ye, and D.~Fan, ``{GDR-HGNN}: A heterogeneous graph neural networks accelerator frontend with graph decoupling and recoupling,'' in \emph{Proceedings of the 61th Annual Design Automation Conference 2024}, ser. DAC '24, 2024.

\bibitem{R-GCN}
M.~Schlichtkrull, T.~N. Kipf, P.~Bloem, R.~v.~d. Berg, I.~Titov, and M.~Welling, ``Modeling relational data with graph convolutional networks,'' in \emph{European semantic web conference}.\hskip 1em plus 0.5em minus 0.4em\relax Springer, 2018, pp. 593--607.

\bibitem{R-GAT}
K.~Wang, W.~Shen, Y.~Yang, X.~Quan, and R.~Wang, ``Relational graph attention network for aspect-based sentiment analysis,'' in \emph{Proceedings of the 58th Annual Meeting of the Association for Computational Linguistics}, 2020, pp. 3229--3238.

\bibitem{ADE-HGNN}
D.~Han, M.~Wu, R.~Xue, M.~Yan, X.~Ye, and D.~Fan, ``{ADE-HGNN}: Accelerating {HGNNs} through attention disparity exploitation,'' in \emph{Euro-Par 2024: Parallel Processing - 30th International Conference on Parallel and Distributed Computing, Proceedings}, ser. Lecture Notes in Computer Science, 2024.

\end{thebibliography}

\end{document}